\documentclass[letter]{aa} 

\usepackage{graphicx}
\usepackage{txfonts}
\def\pd{P_d}

\def\thpc{\theta_{\rm pc}}
\def\thcr{\theta_{\rm cr}}

\def\nct#1{\nocite{#1}}

\def\aap{A\&A}

\def\apjs{ApJS}

\def\lap{\hbox{\hspace{4.3mm}}
         \raise1.5pt \vbox{\moveleft9pt\hbox{$<$}}
         \lower2.5pt \vbox{\moveleft9pt\hbox{$\sim$ }}
         \hbox{\hskip 0.02mm}}
\def\gap{\hbox{\hspace{4.3mm}}
         \raise1.5pt \vbox{\moveleft9pt\hbox{$>$}}
         \lower2.5pt \vbox{\moveleft9pt\hbox{$\sim$ }}
         \hbox{\hskip 0.02mm}}

\def\lap{\hbox{\hspace{4.3mm}}
         \raise1.5pt \vbox{\moveleft9pt\hbox{$<$}}
         \lower1.5pt \vbox{\moveleft9pt\hbox{$\sim$ }}
         \hbox{\hskip 0.02mm}}

\begin{document}

   \title{Geometry of radio pulsar signals:  The origin \\of pulsation
modes and nulling}


   \author{J. Dyks
          }

   \institute{Nicolaus Copernicus Astronomical Center, Polish Academy of Sciences, Rabia\'nska 8, 87-100, Toru\'n,
Poland\\
              \email{jinx@ncac.torun.pl}
             }

\date{Received 2021 May 4; accepted 2021 August 11}


  \abstract
   {
Radio pulsars exhibit an enormous diversity of single pulse behaviour that
involves sudden changes in pulsation mode and nulling occurring on timescales
of tens or hundreds of spin periods. The pulsations appear both chaotic and
quasi-regular, which has hampered their interpretation for decades. Here I show that the pseudo-chaotic 
complexity of single pulses is caused by the viewing of a 
relatively simple radio beam that has a sector structure traceable to the
magnetospheric charge distribution. The slow $\vec E \times \vec B$ 
drift of the sector beam, when
sampled by the line of sight, produces the classical drift-period-folded 
patterns known from observations. The drifting azimuthal zones of
the beam produce the changes in pulsation modes and both the intermodal and
sporadic  nulling at timescales of beating between the drift and the star spin. The axially symmetric conal
beams are thus a superficial geometric illusion, and the standard
carousel model of pulsar radio beams 
does not apply.  
The beam suggests a particle flow structure that involves inward motions with possible
inward emission.} 



   \maketitle
%

\section{Introduction}

Radio pulsars tend to suddenly change their pulsation modes, which involves
a change in sub-pulse drift and a change in average profile on a timescale of tens or
hundreds of star spin periods, $P$ (Weltevrede 2016; Srostlik \& Rankin 2005). 
\nct{sr2005, w2016}
In some modes, the pulse shapes are completely changed with every period (Hankins \&
Wright 1980), and\nct{hw80} 
in other cases they maintain similarity 
throughout the mode. 
In PSR B1237$+$25, with a modulation period, $\pd$, of $\sim2.8P$, the pulses appear
in a repeated sequence with a pulse component  
on the left side, then on the right side, and finally in the middle of
the profile. 
In this same pulsar (Srostlik \& Rankin 2005),\nct{sr2005} 
the cessation of emission (nulling) has been observed
to occur preferentially between the distinct pulsation modes.  In several
objects (e.g.~B1919$+$21; Pr\'oszy\'nski \& Wolszczan 1986)\nct{pw86} 
the flux modulations occur at fixed pulse longitude
$\Phi$ with no drift; however, peculiar $180^\circ$ jumps in modulation
phase are observed (also in B0320$+$39; Edwards et al.~2003).\nct{esv03} In several other objects
(e.g.~PSR B1918$+$19; Rankin et al.~2013) \nct{rwb13} the drifting sub-pulses are observed only
in the profile interior, whereas 
the fixed-$\Phi$ modulations are limited to the peripheral components. The
trend for single pulse emission to move from one side of the profile to the other 
is evident.  One side is then dominating, which creates pseudo-symmetric 
(antisymmetric) profiles with core and conal components, albeit
with only the left or right part of the profile filled in   
(J2145$-$0750; Stairs et al.~1999; Dai et al.~2015). \nct{stc99, dhm15} 
In this paper a radio beam is presented that explains these phenomena.

The radio beam has long been suspected to consist of two nested hollow cones  
(Rankin 1983), \nct{ran83} the radii of which have been derived
from profile modelling (Rankin 1990; 1993). \nct{ran90, ran93}
The only magnetospheric structure suggested for the observed cone size ratio  
was that of critical magnetic lines, located at 
$\theta_{cr}=(2/3)^{3/4}\thpc=0.74\thpc$, where $\thpc$ is the angular polar cap radius 
(Wright 2003).\nct{wri03}

Although the $\vec E \times \vec B$ 
drift has long been considered as the origin of simple sub-pulse drift
(Ruderman \& Sutherland 1975; van Leeuwen et al.~2003; Maan 2019; McSweeney
et al.~2019),\nct{rs75, maa2019, vanl2003, swee2019} the ansatz of the axially symmetric carousel of
sparks made it difficult to interpret all the other phenomena. The pulsations themselves, and thus the modulations, could be interpreted as time variability 
(Clemens \& Rosen 2008),\nct{cr2008} relativistically outflowing layers (Kirk et al.~2002), \nct{ksg02} 
or the laterally moving substructure of
the emission region. Time variability in the spectral space could also produce
the pulsations with the emitted radio spectrum moving across the telescope
band. 

\begin{figure}
\begin{center}
\includegraphics[width=0.47\textwidth]{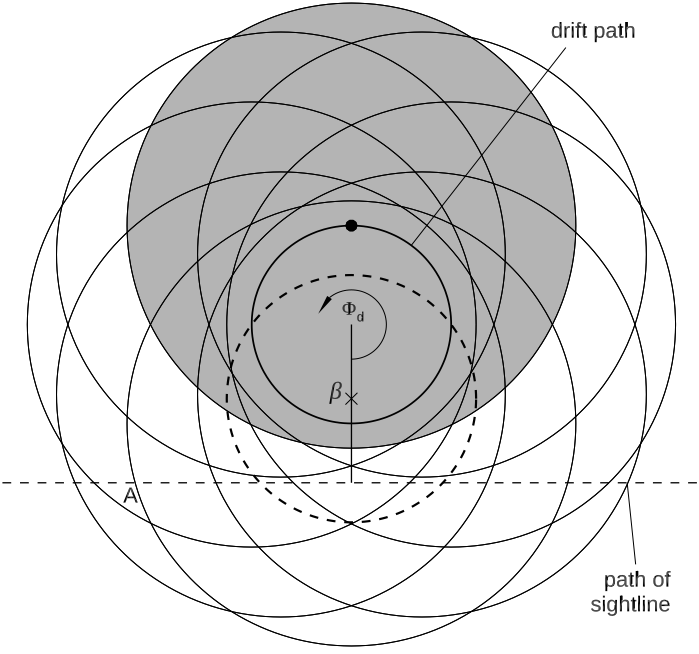}
\end{center}
\caption{Drift of a wide emission region (grey) around the dipole axis. 
Far from the profile
centre (point A), the region is visible within a limited interval of drift
cycle, and strong modulations appear. The extra beam (dashed) on the opposite
side of the dipole axis is required to explain the jump by half of the modulation
cycle.  
}
\label{visi}
\end{figure}

In a parallel paper on B1919$+$21 (Dyks, van Straten, Primak et al.,
in preparation), a successful model of modulated pulsar polarization is set forth. 
The model supports  
the drift interpretation of pulse modulations, 
suggesting that they must correspond to the mapping of the observed flux straight on the
lateral structure of the drifting beam. 
The basic polarization model involves a single radio beam of radius $\rho$  
(grey circle in Fig.~\ref{visi}) that is  
drifting around the dipole axis and is probed by the line of sight once per
spin period, $P$.  This is the starting point from which a more realistic radio beam is
invoked below.

\begin{figure}
\begin{center}
\includegraphics[width=0.5\textwidth]{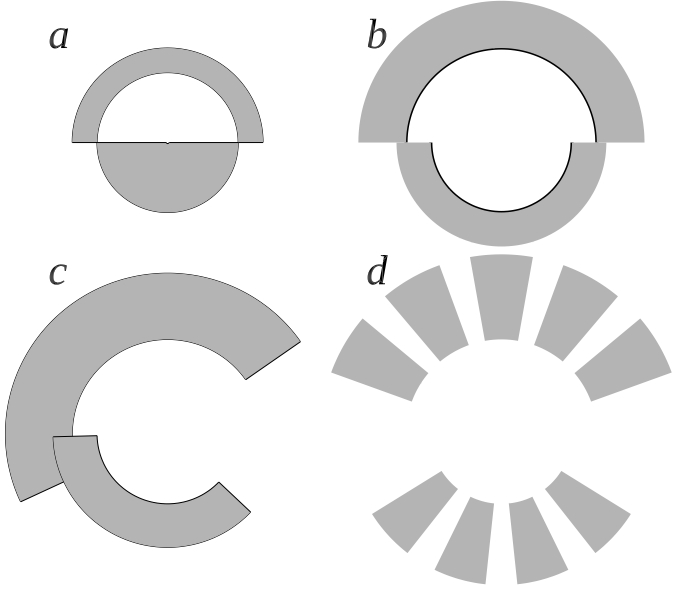}
\end{center}
\caption{Basic beam pattern ({\bf a}) with possible modifications. 
{\bf (a):}
Fixed-altitude emission at all magnetic co-latitudes. The outer and inner
circles correspond to the last open and critical B-field lines, respectively.
{\bf (b):} Emission from
the last open and critical lines within a finite altitude range.  {\bf (c):}
 Arbitrary geometry of sub-beams. {\bf (d):} Radially extended azimuthal structure as invoked from PSR 
B0826$-$34.}
\label{possi}
\end{figure}

\begin{figure}
\begin{center}
\includegraphics[width=0.5\textwidth]{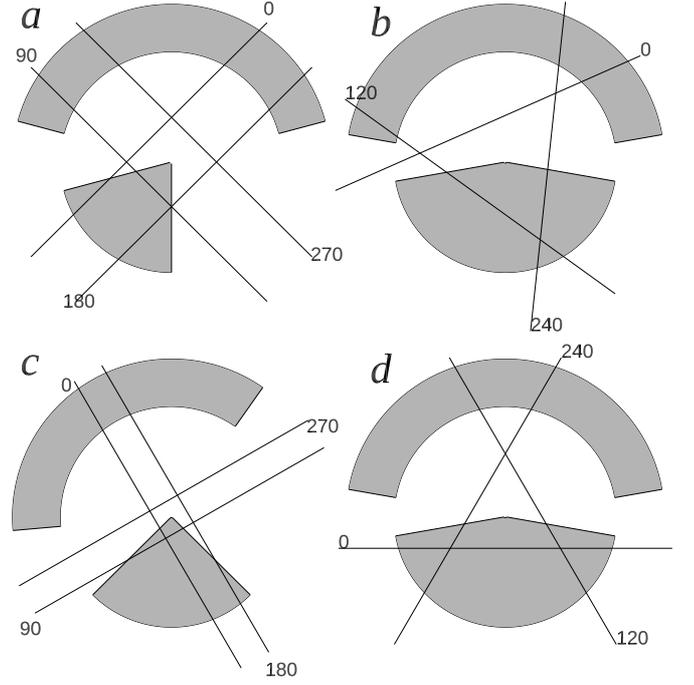}
\end{center}
\caption{Head-on view of a generic radio pulsar beam (grey) invoked from pulse modulation
properties. Thin solid sections present the paths of sightline for low-order resonant
drift ($nP=m\pd$) and different impact angles, 
$\beta^\prime$. Numbers are consecutive azimuths in beam frame. 
The left-side cases show the asymmetry invoked for several pulsars.
}
\label{beam}
\end{figure}

\nct{esa2005}

\section{Radio pulsar beam geometry}

The modulation properties of
PSR B1919$+$21 and B1237$+$25 (see Figs.~3 and 2 in Pr\'oszy\'nski \& Wolszczan
1986) \nct{pw86} provide fair guidance for the beam structure. 
In the profile periphery, the drifting beam of Fig.~\ref{visi} is visible 
only through a limited part of drift period, $\pd$, which corresponds to  
the grey circle passing through point A in Fig.~\ref{visi}. However,  
closer to the profile centre (thus closer to the dipole axis), this grey beam must be missing 
(since a peak flux is replaced with a minimum in the modulation pattern), and
another beam must be located at a half modulation cycle later. Half of a modulation
phase 
corresponds to $\pd/2$; therefore,~the near-axis beam must be located on the other side of
the dipole axis 
(the dashed circle in Fig.~\ref{visi}). We must have an outer sub-beam and an inner sub-beam, each on
the opposite side of the polar tube. In fact, the polar tube is predicted to be divided into the
outer and inner part by the critical magnetic field lines defined by the
Goldreich-Julian charge density. This leads to the geometry shown in
Fig.~\ref{possi}a. If the altitude scaling is ignored ($\propto
r^{1/2}$), the grey outer arc has a radial width of
$1.5\thpc-1.5\thcr$. 

The illustrated structure presents an average beam and reflects only crucial
features.\ For example, the partial overlap of the sub-beams in co-latitude $\theta$, likely resulting from 
magnetic line curvature, is ignored; however, this aspect is not important for understanding the key modulation properties. 
Another beam of Fig.~\ref{possi}b shows the radially extended case,
in which the radio emission is limited to the critical and last open field
lines. The beam consists of two half-rings that result from the flaring of B-field
lines. In Fig.~\ref{beam} the azimuthal extent has been limited to allow for
nulling in the case of near-central viewing. The beam is shown in the co-drifting
frame, that is, the beam rotates around its centre (and dipole axis) at period $\pd$. 
In the case of resonant drift ($nP=m\pd$), the beam is passed along a fixed set of
viewing paths that are shown in the figure for low resonance orders. 
The paths keep a constant impact
angle, $\beta$ (angular distance from the drift or beam 
centre) 
and are marked with numbers that give their steadily increasing
angle of orientation. The passage typically occurs either only from the numbered tip to the tip
without the number or vice versa. To pass in both directions requires
special conditions ($\beta=0$ and, for example, $\pd=n2P$). 

\subsection{Origin of pulsation modes and nulling}

In the case of non-resonant sampling, the viewing paths may slowly rotate with
respect to the beam (while keeping the fixed $\beta\ne0$). When the viewing path 
rotates through the horizontal orientation orthogonal to the beam symmetry plane
(which is vertical in Fig~\ref{beam}), the
pulsation mode is changed. When $\beta$ is small and the sub-beams do not span $180^\circ$ in
azimuth, a null will be observed between pulsation modes. This is where
nulls are indeed often
observed in pulsars (e.g.~Fig. 5 in Srostlik \& Rankin 2005). \nct{sr2005}   

For $\beta=0$ and a uniform motion of viewing paths through the beam, 
only the nulling and two pulsation
modes 
would be possible: `outer sub-beam--inner sub-beam' and `inner--outer'. 
However, for $\beta\ne 0$ at least four pulsation modes are possible:
`outer--inner', `inner--outer', `outer--outer', and `inner' or `core'. 
Moreover, in the case of the resonant drift, the beam 
is sampled at a fixed set of paths with specific orientations. Then, the resonant set of viewing paths
 can define a number of profile shapes that are repeatedly observed in
consecutive order. For example, in the top-right
case in Fig.~\ref{beam}, three different pulse shapes will be observed.  
In the case of disturbance of the drift motion (e.g.~the charge flow stop and
revival), the relative phase of the viewed paths and the beam orientation can
change, which can also change the pulsation mode. 
 For peripheral viewing ($\beta$ larger than the inner sub-beam radius),
only the outer sub-beam can be
detectable, which can lead to a nulling behaviour of a different type than for the central sightline passage.

\subsection{Interpretation of PSR B1237$+$25}

PSR B1237$+$25 exhibits a sequence of pulsation modes that are 
consistent with the not-quite-resonant rotation of the viewing path 
with respect to the beam in Fig.~\ref{beam}. In one of the modes, the components appear first at the
leading side (LS) of the profile, then at the trailing side (TS), and finally in the
centre. The sequence has been interpreted as a spiral motion (the S-burst in Hankins \& Wright 1980). 
\nct{hw80} Since $\pd\sim 2.8P$ is close to $3P$ in B1237$+$25, the burst can be 
tentatively interpreted with 
Fig.~\ref{beam}d. Let the first passage correspond
to the viewing path marked $120$ (with the sightline moving towards the number). 
This produces the LS outer conal component 
and the inner conal component 
on the TS (see the top pulse in Fig.~2 of Hankins \&
Wright). In the next pulse, the path at $240$ is followed (again, towards the
end with the number), producing the
LS inner conal component and the TS outer conal component. In the third
period the inner cone-core complex is observed along path 0. 
Sometimes all five components are observed together (see Fig.~2 in Hankins \& Wright), 
which suggests that the inner and outer sub-beams temporarily and partially
overlap in azimuth 
 (as on the left side in Fig.~\ref{possi}c). 
The model can explain several other effects observed in
PSR B1237$+$25, for example~the tendency for the appearance of component 1 together with 4, and of
component 2 together with 5. The abnormal mode from Srostlik \& Rankin 
appears when the viewing moments pick up the LS of the inner cone-core complex. The
modes are changed when the viewing path traverses through the blank space
between the inner and outer sub-beams, which can lead to the intermodal nulling 
when this drift phase happens to be sampled. 
The modes can also change with partial nulls or 
without nulls. The latter may happen for several reasons,
the most important being the asymmetry of the inner sub-beam shown in 
Fig.~\ref{beam}a: While passing through the vertical orientation, the outer
half-cone is always visible, but the sightline 
gains (or loses) the view of the inner beam without a null.

The conclusion is that the overall modulation properties of PSR
B1237$+$25, along with several details, 
can be understood as the result of the sampling of the beam shown in
Fig.~\ref{beam}. The spiral is not required (though two spiraling arcs could
be inscribed into the grey sub-beams, sharing the basic properties of
the zonal beam: 
imagine the thick solid arcs as having different $\theta$ at each
end).

\begin{figure}
\begin{center}
\includegraphics[width=0.5\textwidth]{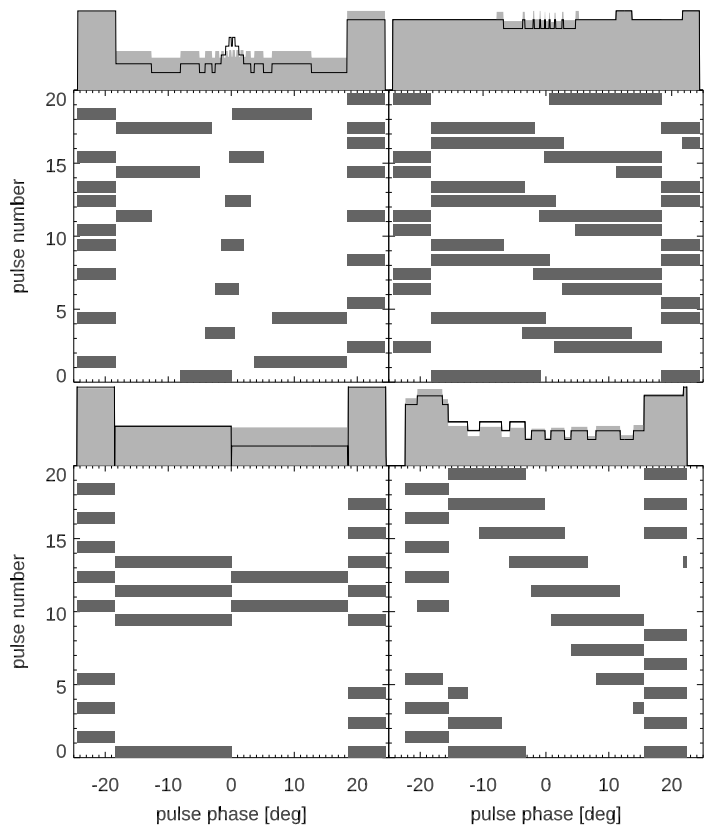}
\end{center}
\caption{Single pulses calculated for the beam types shown in grey at the top of 
Fig.~\ref{beam}. The figure shows several classical single pulse effects: left-right-middle
modulation with core pulses (top left), jump by half of the modulation
phase, with sporadic nulls in pulses 2 and 19 (top right), pulsation mode change at
intermodal nulling (bottom left), and interior drift of sub-pulses flanked by
fixed-longitude flux modulations (bottom right). The corresponding partial profiles are shown
on top with a line and the average profiles for 2000 pulses in grey.  
Parameter values are (from left to right, top to bottom) $\pd=2.9, 4.2, 2.8, 1.9$, $\beta^\prime=-0.1, 0.1,
0, 0.4$, $(\phi_{\rm out}^{\rm min}, \phi_{\rm out}^{\rm max})
=(-70^\circ,70^\circ)$ everywhere, $ \phi_{\rm in}^{\rm min}=110^\circ$
everywhere, and $\phi_{\rm in}^{\rm max}=180^\circ$, except for the top right, where $\phi_{\rm in}^{\rm
max}=250^\circ$; thus, only the top-right panel is for the symmetric  beam (Fig.~\ref{beam}, top right).  
}
\label{singles}
\end{figure}

\subsection{Examples of modelled pulsation modes}

A simple code samples the beams of Fig.~\ref{beam} at rotation angles
\begin{equation} 
\Phi_{\rm cut}=(2\pi/\pd)(i_p + W_j/360^\circ) + \phi_{\rm ct0} \ \rm [rad]
,\end{equation}
where $\pd$ is in $P$, $i_p$ is the pulse number, $\phi_{\rm ct0}$ (in radians) is a constant reference drift
phase, and $W_j$ (in degrees) is the vector of longitudes (of pulse width
length). The viewing path $(W_j,\beta)$ is then rotated by $\Phi_{\rm cut}$ while maintaining
the fixed impact angle $\beta=\beta^\prime1.5 \thpc$. The intra-beam polar coordinates
$(\phi_b,\theta_b)$ are then calculated for each $W_j$, and the conditions to
fall into the grey area of Fig.~\ref{beam} are applied, which includes the azimuthal
span of the sub-beams: $\Delta\phi_{\rm out}$ and $\Delta\phi_{\rm in}$ or,
more generally, the beam limiting azimuths $\phi_{\rm out}^{\rm min}$, $\phi_{\rm out}^{\rm
max}$, $\phi_{\rm in}^{\rm min}$, and $\phi_{\rm in}^{\rm max}$. The
size ratio of the radial zones is here assumed to be fixed by $\thcr/\thpc$ but in general
involves $\theta_{\rm out}^{\rm min}$, $\theta_{\rm out}^{\rm
max}$, $\theta_{\rm in}^{\rm min}$, and $\theta_{\rm in}^{\rm max}$. Asymmetries
and overlaps of the sub-beams are controlled by the last eight parameters.

Example code output 
is shown in Fig.~\ref{singles}. The top-right panel is for the symmetric beam
of Fig.~\ref{beam}b, whereas the other cases are for the asymmetric beam
of Fig.~\ref{beam}a.

The top-left case ($\pd=2.9P$, $\beta^\prime=-0.1$) reveals the
left-right-middle sequence that so misleadingly resembles 
the spiral in B1237$+$25 (Hankins \&
Wright 1980). We note that there is a core component despite no core being in the beam.
In the top-right case ($\pd=4.2P$, $\beta^\prime=0.1$)  the symmetric beam results 
in the half-phase modulation jumps, such as those observed  in B1919$+$21. 
We note the sporadic single pulse nulls at pulses 2 and 19.
The bottom-left case ($\pd=2.8P$, $\beta^\prime=0.0$) shows clear changes in pulse modes 
separated by intermodal
nulls, such as those observed in B1237$+$25 (Fig.~5 in Srostlik and Rankin). The bottom-right
case shows a clear interior drift flanked by the fixed-longitude modulation,
which is observed in several objects (e.g.~B1918$+$19).  

The pulsations have an obvious tendency to take on a quasi-conal look with
several asymmetries, which are mostly related to the sub-beam being located on each
side of the profile. Interestingly, pulses from an apparently inner cone are
sometimes visible in several patterns, despite only one half-cone and a
birthday-cake-like wedge being in the beam. They arise from cutting the corners of the sub-beams. 
The core component appears in many simulations even without any brightening at
the beam centre and results from cutting through the central tip of the inner sub-beam.

 To obtain the main types of behaviour shown in
Fig.~\ref{singles}, in particular the nulling separating core-dominated and
cone-dominated pulsation modes, the beam must be asymmetric
(Fig.~\ref{beam}a).  
Then, over some limiting azimuths
the outer sub-beam may be sampled without the inner sub-beam visible, with mode
transition occurring either at nulls or without nulls (bottom-left case in
Fig.~\ref{singles}). Though not shown, the model also produces the antisymmetric
profiles with only one side bright. The zonal structure of the beam implies
correlations between a main pulse (MP) and an inter-pulse (IP), as well as anti-correlations (Weltevrede et
al.~2012). \nct{wwj12}

It is found that several classical types of single pulse behaviour, reported in
a multitude of observations, can be interpreted with the sector (zonal) beam of
Fig.~\ref{beam}. 

 To properly model the average profiles, a more
realistic emissivity profile (non-uniform) must still be implemented. 
However, calculations made so far make it clear that 
the question of whether the average profile is
brightest at the edges or at the centre 
clearly depends
on the ratio $R_\phi=\Delta\phi_{\rm out}/\Delta\phi_{\rm in}$ and not only on the
impact parameter. 
The average profiles shown in grey in Fig.~\ref{singles} correspond to
the uniform and equal emissivity within the sub-beam zones. If the drift is not resonant, 
the strength ratio of central to peripheral 
components roughly corresponds to the ratio of the azimuthal extent of the
sub-beams, $R_\phi=\Delta\phi_{\rm in}/\Delta\phi_{\rm out}$. The
top-right case in Fig.~\ref{singles} has a boxy shape since $R_\phi=1$,
whereas in the other cases $R_\phi=0.5$, hence, the average profiles have a bright periphery
and correspond to a double type (or perhaps a blended multiple with bright edges). 
Beyond the proposed model, the diversity of average profiles may furthermore be increased by the dependence of
magnetospheric current flow on dipole tilt and by polarization-mode-sensitive effects 
(absorption, scattering, refraction). 

The fractional azimuthal extent also affects the statistics of nulling. 
In the case of non-resonant drift and $\beta=0$, the fraction of nulls
corresponds to the fraction of azimuths that are radio quiet on both sides
of the beam (at $\phi$ and $\phi+\pi$). For
$\beta\ne0$ the null fraction is no longer trivial to estimate because it
depends on both the azimuthal and radial ($\Delta\theta$) extent of the sub-beams.
For example, when the beam becomes narrow in azimuth ($\Delta\phi \ll
\Delta\theta$), it is $\beta$ and $\Delta\theta$ that determine the null
fraction.

Furthermore, the model implies two types of nulling, central and peripheral. 
The second type 
appears when our line of sight is just grazing the beam, with
$\theta_{\rm in}^{\rm max}< \beta < \theta_{\rm out}^{\rm max}$. 
In this case the nulling fraction essentially corresponds to 
$(2\pi-\Delta\phi_{\rm out})/2\pi$. 
A sample of nulling pulsars may thus contain objects that
null in either way (central or peripheral).

If the drift period is resonant, the null
statistics and average profiles are affected by the ratio $P_d/P$. The null
fraction and the profile shapes 
then depend on the absolute
phase of drift with respect to star spin phase.  Depending on the drift
stability, both periodic and non-periodic nulls are possible (Basu et
al.~2020). \nct{bmm20} 

The proposed geometry implies that, generally, the null fraction 
increases when the azimuthal extent of the sub-beams decreases. 
Space-charge limited flow models (discussed in the next section)
predict that accelerating voltage decreases with distance from the main meridian ($\vec \Omega$-$\vec \mu$
plane). In older pulsars (closer to the death line) the pair production may thus
be possible only close to the $\vec \Omega$-$\vec \mu$ plane. In such objects
the sub-beams may then have a smaller azimuthal extent.  This is in line with
the finding, in  Wang et al.~(2007), \nct{wmj07}  
that older pulsars tend to have larger null fractions; that paper 
also concludes that `nulling and mode changing are
different manifestations of the same phenomenon', 
which is consistent with the model proposed here.  

It is unclear if the model applies to the extremely long nulling observed in
state switching pulsars (Kramer et al.~2006; Young et al.~2015; Stairs et al
2019)\nct{klo06, yws15, slk19}
because this requires special conditions: For the long
nulls, the drift must be essentially resonant or extremely slow. Moreover,
the spin down would have to depend on the drift phase.

\subsection{Fine beam structure: Lesson from B0826$-$34}

The observation of several sub-pulses in a single pulse implies that, at
least for some objects, the sub-beams must be azimuthally structured (if the
time modulation of the sub-beams is excluded).   
PSR B0826$-$34 offers insight here because its
beam is nearly parallel to the rotation axis and the pulsar is viewed at a small
viewing angle 
(Esamdin et al.~2005; Gupta et al.~2004).\nct{esa2005, ggks04} The low-flux minima
separating the MP from the IP in this object have been
interpreted as evidence for two nested carousels separated in
co-latitude (Fig.~10 in Esamdin et
al.). However, according to the generic beam of Fig.~\ref{beam}, 
the IP must correspond to the inner-sub-beam wedge and the MP to the outer half-cone sub-beam. The observed MP and IP are
thus separated 
by the blank space that separates the sub-beams in magnetic azimuth.
To account for the observed sub-pulses, the brightest parts of the beam 
extend radially from the dipole axis, as shown in Fig.~\ref{possi}d.  
There is just the radial (spoke-like) structure with the break in the
azimuth (instead of two nested cones with the break in co-latitude). 
The data on B0826$-$34 (Fig.~1 in Esamdin et al.) also imply that the inner sub-beam has
spectral properties that are quite different from those of the outer cone. This is
another factor that affects the average profile shape.  
The back-and-forth
motion of components suggests the sub-beams may wiggle back and forth instead
of following the monotonic $\vec E \times \vec B$ rotation.

\section{Structure of particle flow}

\begin{figure}
\begin{center}
\includegraphics[width=0.46\textwidth]{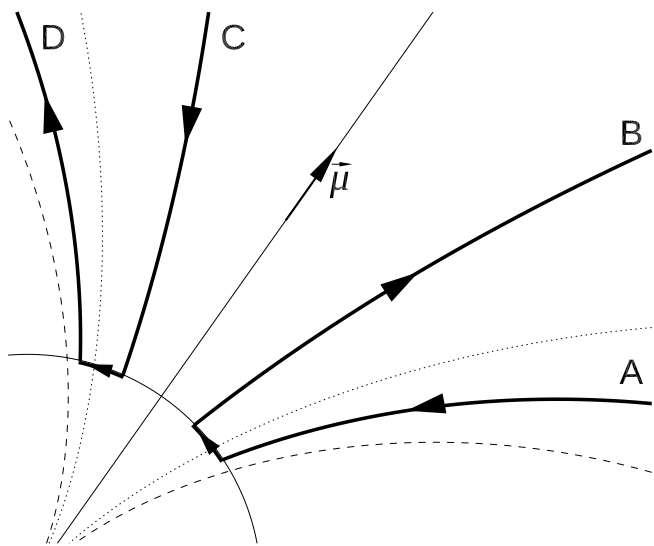}
\end{center}
\caption{Particle flow pattern suggested by the beam in Fig.~\ref{beam}. A reversed flow of emitting particles 
(with all arrows reversed) is also possible. The last open lines are dashed, the
critical lines are dotted, and $\vec \mu$ is the dipole magnetic moment.
}
\label{bfield}
\end{figure}

Assuming that charges within the polar tube are generally capable of
emitting detectable radio emission, the blank parts of the invoked beam
(Fig.~\ref{beam}) can be interpreted as regions with particles flowing away
from the observer. This leads to the particle flow pattern shown in
Fig.~\ref{bfield}. 

The inward flow in regions A and C turns numerous exotic ideas
into reasonable possibilities. If downward flow A is radiating, the
radiation may be obscured at flow B, it can be scattered there, or
it can be mirror-style reflected. It may also pass through to the other side, 
forming the off-pulse quasi-isotropic radiation. 
Extremely wide pulse components with strange absorption features (not
only double notches) have been observed, for example~in PSR B0950$+$08 and B1929$+$10 
(Rankin \& Rathnasree 1997; McLaughlin \& Rankin 2004),\nct{rr97, mr04}
although not in
other pulsars (e.g.~Vela, Kramer et al.~2002). \nct{kjv02}
The asymmetry in
Fig.~\ref{bfield} seems consistent with the non-universal character of this
phenomenon.

When different magnetic azimuths are considered, the emission
from flow A is always directed towards the near-axial region, 
which leads to interesting ray focusing geometry.
Moreover, it is now possible to assume that the radio signal is produced at high altitudes 
in a weaker magnetic field 
but
is emitted inwards, where it is reflected or scattered outwardly.
Similar bidirectional effects are expected for flow CD; however, since flow C is more vertical, 
its radiation is more likely to be reflected outwards by the star surface or a central plasma region.


The loops of flow in Fig.~\ref{bfield}
(such as AB between the periphery and centre of the polar tube)
are generally consistent with expectations (Fig.~3 in Mestel \& Shibata 1994). \nct{ms94}
However, the antisymmetry around $\vec \mu$ is yet to be understood. 
The local excess of charge density depends on whether local B-field lines bend 
towards the closest rotational pole or towards the equator. Therefore, the accelerating electric field $E_\parallel$ 
is opposite in two semicircles of a polar cap (Arons \& Scharlemann 1979; 
Mestel \& Shibata 1994). \nct{as79, ms94}
When the semicircles are superposed on a ring in a polar tube, 
the pattern of Fig.~\ref{possi}a can be expected.  
While referring to the semicircular acceleration region, Beskin (2010, p.~109) \nct{bes2010}
states that 
`accordingly, the radiation emissivity pattern should also have the form of a
semicircle. However, this conclusion contradicts the observational data'.
As shown above, the semicircle (and a half-ring) can be invoked from the
checkerboard-like 
modulation pattern. Still, 
the semicircle of, say, poleward-bending B-field lines does not drift
around the dipole axis. On the other hand, if the lateral drift is attributed to the
outflowing plasma (rather than to the pattern of $E_\parallel$), it may be difficult to
maintain the drifting sub-beams because the relativistic particles
escape the magnetosphere in about $1/6$ of $P$.   
Finally, the ring of critical field lines is
expected for small or mild dipole inclinations and depends on the outer gap
activity. Instead, it is possible that the outer half-ring of the invoked beam 
corresponds to the return current sheet (see Fig.~1
by X.~Bai in Timokhin \& Arons 2013). \nct{ta13}

\section{Conclusions}

The radio pulsar beam is sector-structured both in
azimuth and magnetic co-latitude and is indeed $\vec E \times \vec B$-drifting around the dipole
axis.
The sub-beam zones are essentially antisymmetric, as implied by the
checkerboard-like modulation patterns observed in PSR B1919$+$21 and B1237$+$25. 
The long timescales of pulse moding and nulling result from the
near resonance of $\pd$ and $P$, which causes a very slow rotation of
viewing path  through the beam. 
 
The established geometry produces pulsations that bear
an obvious and striking resemblance to the single pulse observations, explain several types 
of typically observed behaviour, and allow for a much larger 
diversity of effects than the axially symmetric carousel. 
Moreover, the zonal beam is in line with the general structure of the magnetospheric charge density distribution. 

Although multiple questions about the beam details 
remain open, the beam is capable of explanations that are beyond the reach 
of the standard carousel model or the surface oscillation model,  
making these models obsolete. However, the results support the cone size origin proposed
 by Wright (2003). \nct{wri03} 
It is clear that it becomes possible to apply the proposed beam to several observed effects and objects, 
which should allow the beam geometry and behaviour  to be better constrained.

\begin{acknowledgements}
This work was supported by the grant 
2017/\-25/\-B/\-ST9/\-00385 of the National Science
Centre, Poland.
\end{acknowledgements}

\bibliographystyle{aa}
\bibliography{listofrefs2}

\end{document}